**Proofs of Vector Identities Using Tensors**

Zaheer Uddin, Intikhab Ulfat

University of Karachi, Pakistan

---

**ABSTRACT:**

The vector algebra and calculus are frequently used in many branches of Physics, for example, classical mechanics, electromagnetic theory, Astrophysics, Spectroscopy, etc. Important vector identities with the help of Levi-Civita symbols and Kronecker delta tensor are proved and presented in this paper. Some of the identities have been proved using Levi-Civita Symbols by other mathematicians and Physicists. The rests are presented for the first time. The derivations are of interest for both graduate and undergraduate students.

## INTRODUCTION

The Levi-Civita tesnor is totally antisymmetric tensor of rank n. The Levi-Civita symbol $\epsilon_{ijk}$ is also called permutation symbol or antisymmetric symbol. It is named after the Italian mathematician and Physicist Tullio Levi-Civita [1-3].

In three dimensions, it the Levi Civita tensor is defined as

$$\epsilon_{ijk} = \begin{cases} 1 & ijk \ are \ cyclic \\ -1 & ijk \ are \ non \ cyclic \\ 0 & else \ where \end{cases}$$

The indices *i, j, and k* run from 1, 2, and 3. There are 27 values of Levi-Civita tensor components, only six of them are non zero. Three of them are positive one and the other three are negative one. The exchange of any two indices revert the sign of the Levi-Civita symbol. Any tensor whose components form ortho-normal basis can be represented with the help of Levi-Civita symbol, such a tensor is also called permutation tensor.

A Kronecker symbol also known as Knronecker delta is defined as

$$\delta_{ij} = \begin{cases} 0 & i \neq j \\ 1 & i = j \end{cases}$$

$\delta_{ij}$ are the matrix elements of the identity matrix [4-6].

The product of two Levi Civita symbols can be given in terms Kronecker deltas.

$$\epsilon_{ijk}\epsilon_{klm} = \delta_{il}\delta_{jm} - \delta_{im}\delta_{jl}$$

The Kronecker delta and Levi-Civita symbols can be used to define scalar and vector product, respectively [5,6].

$$A \bullet B = A_1B_1 + A_2B_2 + A_3B_3 = \delta_{ij}A_iB_j$$

$$\vec{A} \times \vec{B} = \begin{vmatrix} i & j & k \\ A_1 & A_2 & A_3 \\ B_1 & B_2 & B_3 \end{vmatrix}$$

$\left(\vec{A}\times\vec{B}\right)_1 = A_2B_3 - A_3B_2$

$\left(\vec{A}\times\vec{B}\right)_i = \epsilon_{ijk}A_jB_k$

The repeated indices indicate a sum over these indices.

The Del operations on scalar and vector field are given by

$(i)\ (\nabla\varphi)_{\mathrm{i}} = \dfrac{\partial\varphi}{\partial x_i}$

$(ii)\ \nabla\bullet\vec{A} = \delta_{ij}\dfrac{\partial}{\partial x_i}A_j$

$(iii)\ \nabla\times\vec{A} = \epsilon_{ijk}\dfrac{\partial}{\partial x_j}A_k$

## APPLICATION TO CLASSICAL MECHANICS

In the following section we have taken examples of vectors identities from classical mechanics and proved them using Kroneker delta and Levi-Civita definitions. These identities can also be proved using Cartesian coordinated, but the proof will take longer as compared to the time taken by these proofs.

$\boldsymbol{Example\ 1.}\nabla\bullet(r^3\vec{r}) = \dfrac{\partial}{\partial x_i}(r^3x_i) = 3r^3 + x_i\left(3r^2\dfrac{x_i}{r}\right) = 3r^3 + 3rx_i^2 = 3r^3 + 3r(r^2) = 6r^3$

$\boldsymbol{Example\ 2.}\nabla\bullet\left(r\nabla\left(\dfrac{1}{r^3}\right)\right) = \dfrac{\partial}{\partial x_i}\left(r\left(\dfrac{\partial}{\partial x_i}\dfrac{1}{r^3}\right)\right) = \dfrac{\partial}{\partial x_i}\left(r\left(-3\dfrac{x_i}{r^5}\right)\right) = -3\dfrac{\partial}{\partial x_i}\left(\dfrac{x_i}{r^4}\right)$

$\nabla\bullet\left(r\nabla^2\left(\frac{1}{r^3}\right)\right) = -3\left(\frac{3r^4-4r^2x_i^2}{r^8}\right) = -3\left(\frac{3r^4-4r^4}{r^8}\right) = 3r^{-4}$

$\boldsymbol{Example\ 3.}\nabla\times\left(\dfrac{\vec{r}}{r^2}\right) = \epsilon_{ijk}\dfrac{\partial}{\partial x_j}\left(\dfrac{x_k}{r}\right) = \epsilon_{ijk}(x_k)\left(\dfrac{-x_j}{r^2}\right) = \dfrac{\vec{r}\times\vec{r}}{r^2} = 0$

$\boldsymbol{Example\ 4.}\nabla^2\left(\nabla\bullet\left(\dfrac{\vec{r}}{r^2}\right)\right) = \dfrac{\partial}{\partial x_i}\left[\dfrac{\partial}{\partial x_i}\left\{\dfrac{\partial}{\partial x_j}\left(\dfrac{x_j}{r^2}\right)\right\}\right] = \dfrac{\partial}{\partial x_i}\left[\dfrac{\partial}{\partial x_i}\left\{\dfrac{3r^2-2r^2}{r^4}\right\}\right]$

$\nabla^2\left(\nabla\bullet\left(\dfrac{\vec{r}}{r^2}\right)\right) = \dfrac{\partial}{\partial x_i}\left[\dfrac{\partial}{\partial x_i}\left\{\dfrac{1}{r^2}\right\}\right] = \dfrac{\partial}{\partial x_i}\left[\dfrac{-2x_i}{r^4}\right] = -2\left(\dfrac{3r^4-4r^4}{r^8}\right) = 2r^{-4}$

$\boldsymbol{Example\ 5.}\nabla^2(ln\ r) = \dfrac{\partial}{\partial x_i}\left\{\dfrac{\partial}{\partial x_i}(ln\ r)\right\} = \dfrac{\partial}{\partial x_i}\left\{\dfrac{x_i}{r^2}\right\} = \dfrac{3r^2-2r^2}{r^4} = \dfrac{1}{r^2}$

$\boldsymbol{Example\ 6.}\nabla\bullet\left(\nabla\times\vec{F}\right) = \dfrac{\partial}{\partial x_i}(\nabla\times F)_i = \dfrac{\partial}{\partial x_i}\left(\epsilon_{ijk}\dfrac{\partial}{\partial x_j}F_k\right) = \dfrac{\partial}{\partial x_i}\left(\epsilon_{ijk}\dfrac{\partial}{\partial x_j}F_k\right)$

$\nabla\bullet\left(\nabla\times\vec{F}\right) = \dfrac{\partial}{\partial x_1}\left(\dfrac{\partial}{\partial x_2}F_3 - \dfrac{\partial}{\partial x_3}F_2\right) + \dfrac{\partial}{\partial x_2}\left(\dfrac{\partial}{\partial x_3}F_1 - \dfrac{\partial}{\partial x_2}F_3\right) + \dfrac{\partial}{\partial x_3}\left(\dfrac{\partial}{\partial x_1}F_2 - \dfrac{\partial}{\partial x_2}F_1\right) = 0$

$\boldsymbol{Example\ 7.} \left(\nabla \times (\vec{A} \times \vec{B})\right)_i = \epsilon_{ijk} \frac{\partial}{\partial x_j} (\vec{A} \times \vec{B})_k = \epsilon_{ijk} \frac{\partial}{\partial x_j} (\epsilon_{klm} A_l B_m)$

$$\left(\nabla \times (\vec{A} \times \vec{B})\right)_i = \epsilon_{ijk} \epsilon_{klm} \frac{\partial}{\partial x_j} (A_l B_m)$$

$$\left(\nabla \times (\vec{A} \times \vec{B})\right)_i = (\delta_{il}\delta_{jm} - \delta_{im}\delta_{jl}) \left(A_l \frac{\partial B_m}{\partial x_j} + B_m \frac{\partial A_l}{\partial x_j}\right)$$

$$= \delta_{il}\delta_{jm} A_l \frac{\partial B_m}{\partial x_j} - \delta_{im}\delta_{jl} A_l \frac{\partial B_m}{\partial x_j} + \delta_{il}\delta_{jm} B_m \frac{\partial A_l}{\partial x_j} - \delta_{im}\delta_{jl} B_m \frac{\partial A_l}{\partial x_j}$$

$$= A_i \frac{\partial B_j}{\partial x_j} - A_j \frac{\partial B_i}{\partial x_j} + B_j \frac{\partial A_i}{\partial x_j} - B_i \frac{\partial A_j}{\partial x_j}$$

$$= A_i (\nabla \bullet \vec{B}) - (\vec{A} \bullet \nabla) B_i + (\vec{B} \bullet \nabla) A_i - B_i (\nabla \bullet \vec{A})$$

$\nabla \times (\vec{A} \times \vec{B}) = \vec{A}(\nabla \bullet \vec{B}) - (\vec{A} \bullet \nabla)\vec{B} + (\vec{B} \bullet \nabla)\vec{A} - \vec{B}(\nabla \bullet \vec{A})$ [7]

$\boldsymbol{Example\ 8.} \left(\vec{A} \times (\nabla \times \vec{B})\right)_i == \epsilon_{ijk} A_j (\nabla \times \vec{B})_k = \epsilon_{ijk} A_j \epsilon_{klm} \frac{\partial}{\partial x_l} B_m = \epsilon_{ijk} \epsilon_{klm} A_j \frac{\partial}{\partial x_l} B_m$

$$\epsilon_{ijk} \epsilon_{klm} = \delta_{il}\delta_{jm} - \delta_{im}\delta_{jl}$$

$$\epsilon_{ijk} \epsilon_{klm} A_j \frac{\partial}{\partial x_l} B_m = \delta_{il}\delta_{jm} A_j \frac{\partial}{\partial x_l} B_m - \delta_{im}\delta_{jl} A_j \frac{\partial}{\partial x_l} B_m$$

$$\epsilon_{ijk} \epsilon_{klm} A_j \frac{\partial}{\partial x_l} B_m = A_j \frac{\partial}{\partial x_i} B_j - A_j \frac{\partial}{\partial x_j} B_i = A_j \frac{\partial}{\partial x_i} B_j - (\vec{A} \bullet \nabla) B_i$$

Similarly

$$\left(\vec{B} \times (\nabla \times \vec{A})\right)_i = B_j \frac{\partial}{\partial x_i} A_j - (\vec{B} \bullet \nabla) A_i$$

Adding both the equations

$$\left(\vec{A} \times (\nabla \times \vec{B})\right)_i + \left(\vec{B} \times (\nabla \times \vec{A})\right)_i = A_j \frac{\partial}{\partial x_i} B_j - (\vec{A} \bullet \nabla) B_i + B_j \frac{\partial}{\partial x_i} A_j - (\vec{B} \bullet \nabla) A_i$$

$$\left(\vec{A} \times (\nabla \times \vec{B})\right)_i + \left(\vec{B} \times (\nabla \times \vec{A})\right)_i + (\vec{A} \bullet \nabla) B_i + (\vec{B} \bullet \nabla) A_i = A_j \frac{\partial}{\partial x_i} B_j + B_j \frac{\partial}{\partial x_i} A_j$$

$$\left(\vec{A} \times (\nabla \times \vec{B})\right)_i + \left(\vec{B} \times (\nabla \times \vec{A})\right)_i + (\vec{A} \bullet \nabla) B_i + (\vec{B} \bullet \nabla) A_i = \frac{\partial}{\partial x_i} (A_j B_j)$$

$$\nabla(\vec{A} \bullet \vec{B}) = \vec{A} \times (\nabla \times \vec{B}) + \vec{B} \times (\nabla \times \vec{A}) + (\vec{A} \bullet \nabla)\vec{B} + (\vec{B} \bullet \nabla)\vec{A}$$

$\boldsymbol{Example\ 9.} \left(\vec{A} \times (\nabla \times \vec{A})\right)_i == \epsilon_{ijk} A_j (\nabla \times \vec{A})_k = \epsilon_{ijk} A_j \epsilon_{klm} \frac{\partial}{\partial x_l} A_m = \epsilon_{ijk} \epsilon_{klm} A_j \frac{\partial}{\partial x_l} A_m$

$$\epsilon_{ijk} \epsilon_{klm} = \delta_{il}\delta_{jm} - \delta_{im}\delta_{jl}$$

$$\epsilon_{ijk}\epsilon_{klm}A_j\frac{\partial}{\partial x_l}B_m = \delta_{il}\delta_{jm}A_j\frac{\partial}{\partial x_l}A_m - \delta_{im}\delta_{jl}A_j\frac{\partial}{\partial x_l}A_m$$

$$\epsilon_{ijk}\epsilon_{klm}A_j\frac{\partial}{\partial x_l}B_m = A_j\frac{\partial}{\partial x_i}A_j - A_j\frac{\partial}{\partial x_j}A_i = A_j\frac{\partial}{\partial x_i}A_j - (\vec{A}\bullet\nabla)A_i$$

$$\left(\vec{A}\times(\nabla\times\vec{A})\right)_i = \frac{1}{2}\left(A_j\frac{\partial}{\partial x_i}A_j + A_j\frac{\partial}{\partial x_i}A_j\right) - (\vec{A}\bullet\nabla)A_i$$

$$\left(\vec{A}\times(\nabla\times\vec{A})\right)_i = \frac{1}{2}\left(\frac{\partial}{\partial x_i}(A_jA_j)\right) - (\vec{A}\bullet\nabla)A_i$$

$$\vec{A}\times(\nabla\times\vec{A}) = \frac{1}{2}\nabla A^2 - (\vec{A}\bullet\nabla)\vec{A}$$

## APPLICATION TO ELECTROMAGNETIC THEORY

Example 10 shows application of Kronecker delta on Gauss's law. Poisson and Laplace equation have been derived from this application [8].

$$\boldsymbol{Example\ 10}.\nabla\bullet\vec{D} = \frac{\partial}{\partial x_i}D_i = \epsilon\frac{\partial}{\partial x_i}\ (\nabla V)_i = \epsilon\frac{\partial}{\partial x_i}\left(\frac{\partial V}{\partial x_i}\right) = \epsilon\frac{\partial^2 V}{\partial x_i^2}$$

$$\nabla\bullet\vec{D} = \epsilon\nabla^2 V$$

$$\nabla^2 V = \frac{\rho}{\epsilon},\quad Poisson\ equation.$$

$$For\ free\ space, \nabla^2 V = 0,\quad Laplace\ equation.$$

The solution of Poisson and Laplace equations which satisfies the boundary condition is the only solution that exists, i.e. the solution is unique. This is known as Uniqueness theorem of electromagnetism [8-10]. In example 11, the uniqueness theorem is proved using the identity, divergence of scalar multiple of a vector. Let $V_1\ and\ V_2$ are tow solutions of Laplace or Poisson equations, both the solution satisfy the same boundary condition.

$$\boldsymbol{Example\ 11}.\nabla\bullet(V\vec{D}) = \frac{\partial}{\partial x_1}(VD_i) = V\frac{\partial}{\partial x_i}(D_i) + D_i\frac{\partial}{\partial x_i}(V) = \nabla\bullet(V\vec{D}) = V(\nabla\bullet\vec{D}) + \vec{D}\bullet\nabla V$$

$$\nabla\bullet(V\vec{D}) = \frac{\partial}{\partial x_i}(VD_i) = V\frac{\partial}{\partial x_i}(D_i) + D_i\frac{\partial}{\partial x_i}(V) = \frac{\partial}{\partial x_i}(VD_i) = V\frac{\partial}{\partial x_i}(\epsilon E_i) + \epsilon E_i\frac{\partial}{\partial x_i}(V)$$

$$\nabla\bullet(V\vec{D}) = \epsilon V\frac{\partial}{\partial x_i}\left(\frac{\partial}{\partial x_i}(V)\right) + \epsilon\frac{\partial}{\partial x_i}(V)\frac{\partial}{\partial x_i}(V) = \epsilon V\nabla^2 V + \epsilon(\nabla V)^2$$

If we take volume integral of the equation the integral on left side would be zero by divergence theorem and the first term is zero for free space (Laplace equation), hence

$$(\nabla V)^2 = 0$$

$$\nabla V = 0$$

This shows that V is constant, hence$V_1 = V_2$, the solution is unique.

$$\boldsymbol{Example\ 12.}\left(\nabla\times(\varphi\vec{J})\right)_i=\epsilon_{ijk}\frac{\partial}{\partial x_j}(\varphi\vec{J}_k)=\epsilon_{ijk}\frac{\partial\varphi}{\partial x_j}(\vec{J}_k)+\varphi\epsilon_{ijk}\frac{\partial}{\partial x_j}(\vec{J}_k)$$
$$=(\nabla\varphi\times\vec{J})_i+\varphi(\nabla\times\vec{J})_i$$

$$\nabla\times(\varphi\vec{J})=(\nabla\varphi)\times\vec{J}+\ \varphi(\nabla\times\vec{J})$$

The ampere Circuital law in point form is given by$\nabla\times\vec{H}=\vec{J}$. This is one of the four Maxwell's equations of electromagnetism. In the following example this equation is proved by the application of Levi-Civita and Kronecker Symbols on vector expression, $\nabla\times(\nabla\times\vec{A})$.

$$\boldsymbol{Example\ 13.}\nabla\times\vec{H}=\nabla\times\frac{\vec{B}}{\mu_o}=\frac{1}{\mu_o}\nabla\times\vec{B}=\frac{1}{\mu_o}\nabla\times(\nabla\times\vec{A})$$

$$\left(\nabla\times(\nabla\times\vec{A})\right)_{\mathrm{i}}==\epsilon_{ijk}\frac{\partial}{\partial x_j}(\nabla\times\vec{A})_k=\epsilon_{ijk}\frac{\partial}{\partial x_j}\epsilon_{klm}\frac{\partial}{\partial x_l}A_m=\epsilon_{ijk}\epsilon_{klm}\frac{\partial}{\partial x_j}\frac{\partial}{\partial x_l}A_m$$

$$\epsilon_{ijk}\epsilon_{klm}=\delta_{il}\delta_{jm}-\delta_{im}\delta_{jl}$$

$$\epsilon_{ijk}\epsilon_{klm}\frac{\partial}{\partial x_j}\frac{\partial}{\partial x_l}A_m=\delta_{il}\delta_{jm}\frac{\partial}{\partial x_j}\frac{\partial}{\partial x_l}A_m-\delta_{im}\delta_{jl}\frac{\partial}{\partial x_j}\frac{\partial}{\partial x_l}A_m$$

$$\epsilon_{ijk}\epsilon_{klm}\frac{\partial}{\partial x_j}\frac{\partial}{\partial x_l}A_m=\frac{\partial}{\partial x_j}\frac{\partial}{\partial x_i}A_j-\frac{\partial}{\partial x_j}\frac{\partial}{\partial x_j}A_i=\frac{\partial}{\partial x_i}(\nabla\bullet\vec{A})-(\nabla^2\vec{A})_i$$

$\nabla\times(\nabla\times\vec{A})=\nabla(\nabla\bullet\vec{A})-\nabla^2\vec{A}$ [11].

$$\nabla\times\vec{H}=\frac{1}{\mu_o}\left(\nabla(\nabla\bullet\vec{A})-\nabla^2\vec{A}\right)$$

For static or dc condition$\nabla\bullet\vec{A}=0$, where $\vec{A}\ is$ vector potential. The Poisson equation is given by $\nabla^2V=\frac{\rho}{\epsilon}$ by analogy we can write, $\nabla^2\vec{A}=-\mu_o\vec{J}$, hence from above equation gives $\nabla\times\vec{H}=\vec{J}$.

## CONCLUSION

The vectors in physics teaching, is an important topic. The vectors are taught both at undergraduate and graduate levels. The vector identities appear complicated in standard vector notations. Here we have provided proofs of these vector identities by an alternate method (by the use of Kronecker and Levi-Civita symbols). Altogether we discussed thirteen examples of vector identities from mechanics and electromagnetism. The advantage of using this method is two-fold. Firstly, students will find that this method is short, simplified and straightforward; secondly, this method requires lesser time to prove the identities. The proofs presented in this paper are new flavor for teaching vectors to physics and mathematics students.